\documentclass{INTERSPEECH2023}

\usepackage{url}
\usepackage{hyperref}  
\usepackage{times}
\usepackage{soul}
\usepackage[utf8]{inputenc}
\usepackage{caption}

\usepackage{amsmath}
\usepackage{amsthm}
\usepackage{booktabs}
\usepackage{algorithm}
\usepackage{algorithmic}
\usepackage[switch]{lineno}
\usepackage{makecell}
\usepackage{adjustbox}
\usepackage{mathtools}
\usepackage{subcaption}
\usepackage{lipsum}
\usepackage{wrapfig}
\usepackage{graphicx}
\usepackage{tablefootnote}

\usepackage{dsfont}

\interspeechcameraready 

\title{Diff-HierVC: Diffusion-based Hierarchical Voice Conversion with Robust Pitch Generation and Masked Prior for Zero-shot Speaker Adaptation}
\name{Ha-Yeong Choi, Sang-Hoon Lee, Seong-Whan Lee$^{\dagger}$\thanks{$^\dagger$Corresponding author}}
\address{
Department of Artificial Intelligence, Korea University, Seoul, Korea
}
\email{\{hayeong, sh\_lee, sw.lee\}@korea.ac.kr}
\begin{document}
\maketitle
\begin{abstract}
Although voice conversion (VC) systems have shown a remarkable ability to transfer voice style, existing methods still have an inaccurate pitch and low speaker adaptation quality. To address these challenges, we introduce Diff-HierVC, a hierarchical VC system based on two diffusion models. We first introduce DiffPitch, which can effectively generate $F_0$ with the target voice style. Subsequently, the generated $F_0$ is fed to DiffVoice to convert the speech with a target voice style. Furthermore, using the source-filter encoder, we disentangle the speech and use the converted Mel-spectrogram as a data-driven prior in DiffVoice to improve the voice style transfer capacity. Finally, by using the masked prior in diffusion models, our model can improve the speaker adaptation quality. Experimental results verify the superiority of our model in pitch generation and voice style transfer performance, and our model also achieves a CER of 0.83\% and EER of 3.29\% in zero-shot VC scenarios.     
\end{abstract}
\vspace{-0.2cm}
\section{Introduction}
\vspace{-0.1cm} 
Voice conversion (VC) tasks typically convert the voice of a source speaker into the voice of a specific target speaker, and the linguistic information of the converted target speaker must be consistent with the source speech. The primary concept of VC is to disentangle the individual components of speech so that each component can be controlled and transformed to the target speaker voice. Recently, the VC system has significantly advanced with deep learning approaches, allowing clarity and naturalness of the converted voice \cite{qian2019autovc,lee2021voicemixer,polyak21_interspeech,choi2021neural,popov2022diffusionbased,lee2022duration}. Moreover, the expansion of the conversion system has enabled effective applications in a variety of fields, such as cross-lingual \cite{zhao2021towards,choi2023dddm} and emotional VC \cite{zhou2020converting,zhou2022emotional}. Despite these advancements, converted voice is still perceived as unnatural owing to mispronunciation in converted speech, and low speaker adaptation performance is still a challenge that requires addressing \cite{huang2022flowcpcvc,yuan2022deid}.

Pitch modeling is essential to achieving speech intelligibility and naturalness in VC and text-to-speech (TTS) tasks. Pitch characteristics are crucial in speaker identity and correct pronunciation \cite{lancucki2021fastpitch}.
\cite{qian2020f0,qian2020unsupervised} train the model using normalized fundamental frequency ($F_0$) to obtain the same mean and variance for all speakers. 
This approach contributes to expressiveness by considering the pitch information. However, $F_0$ is not entirely separated from the speaker style, so that it still causes perceptual unnaturalness in the conversion. SR \cite{polyak21_interspeech} uses a vector quantized variational auto-encoder (VQ-VAE) to learn a speaker-irrelevant pitch representation. Although speaker-irrelevant pitch can be extracted, mispronunciation occurs due to the loss of pitch information during vector quantization. In addition, it is difficult to precisely predict the pitch of a voice with a high degree of expressiveness.  
To address the above problem, we propose Diff-HierVC, a novel diffusion-based hierarchical VC system. Diff-HierVC consists of DiffPitch and DiffVoice, which hierarchically convert the voice style from disentangled speech representations. DiffPitch generates the pitch information of the target speaker during the inference step, and DiffVoice constructs a high-quality Mel-spectrogram utilizing the generated pitch information and the source-filter representation according to the source-filter theory. We found that a hierarchical VC architecture is an effective structure for decoupling speech components and generating the converted speech. Moreover, using the data-driven prior, we improved the conversion performance by regulating the inception of the denoising process of the diffusion model. 
Furthermore, we introduce a masked prior that allows the diffusion model to consider the context and condition for better generalization ability and robust training. The experimental results denote that pitch and voice modeling have considerable effects.  
Our main contributions are summarized as follows:
\begin{itemize}
    \item We propose Diff-HierVC, a diffusion-based hierarchical VC system with robust pitch generation and masked prior for expressive zero-shot voice style transfer.
    \item To the best of our knowledge, this is the first study to utilize the diffusion process to generated $F_0$. We demonstrated that using the generated $F_0$ by the denoising diffusion process rather than conventional pitch modeling methods resulted in more accurate pronunciation and natural intonation of the converted voice.
    \item The experimental results reveal that Diff-HierVC achieves a significantly improved zero-shot style transfer in various conversion scenarios with cross-lingual and expressive real-world speech dataset. Our demos are available at \url{https://diff-hiervc.github.io/}.
\end{itemize} 
\vspace{-0.35cm}
\section{Background: diffusion models}
\vspace{-0.1cm} 
Diffusion models have shown extraordinary performance in generative tasks in various domains, such as images, videos, and audio, and have recently achieved considerable success in multi-modal tasks \cite{kawar2023imagic,ruan2023mm}. Specifically, in speech, the diffusion model is utilized in applications such as audio generation \cite{kong2021diffwave,chen2021wavegrad}, speech enhancement \cite{welker22_interspeech}, and TTS synthesis \cite{popov2021grad,huang2022fastdiff}. 
The fundamental concept underlying the stochastic differential equation (SDE)-based continuous-time diffusion process \cite{song2020score} is to train an estimator that repeatedly removes noise by estimating log-density gradient of data and generates samples with an iterative denoising process via SDE. The SDE-based diffusion model was also applied to VC task using a maximum likelihood (ML)-SDE solver \cite{popov2022diffusionbased} for fast sampling.

\begin{figure*}[!t]
  \centering
\centerline{\includegraphics[width=0.96\textwidth]{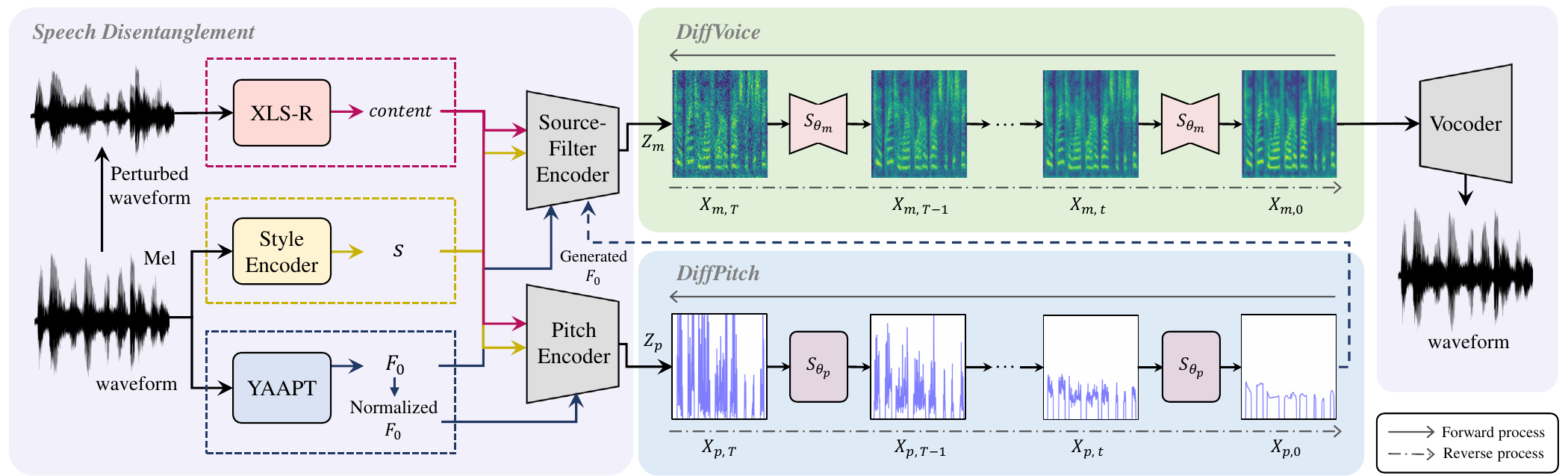}}
\vspace{-0.2cm}\caption{Overall framework}\vspace{-0.5cm}
\label{framework}
\end{figure*}
\newpage
\section{Diff-HierVC} 
\subsection{Speech disentanglement} 
As illustrated in Figure \ref{framework}, we first analyze speech into representations of content, pitch, and style: 
(1) Data perturbation \cite{choi2021neural} is applied to the input waveform to eliminate content-irrelevant information. Subsequently, we extract the content features from the intermediate layer representation of XLS-R \cite{babu22_interspeech}, a pre-trained self-supervised model using a large-scale cross-lingual speech dataset.
(2) We utilize a style encoder \cite{min2021meta} to extract the voice style, which is the speaker style representation from the Mel-spectrogram. The style embedding serves as a guide for both the content encoder and pitch encoder.
(3) We extract a fundamental frequency ($F_0$) using the YAAPT algorithm \cite{kasi2002yet} with a 4$\times$ high-resolution higher than Mel-spectrogram for precise pitch extraction. The content encoder receives $\log(F_0+1)$, and the pitch encoder takes the normalized $F_0$ as the mean and variance of the source speaker's $F_0$.
 
\subsection{Hierarchical VC}
For hierarchical VC, we introduce a two-stage diffusion models, DiffPitch and DiffVoice. DiffPitch initially converts the $F_0$ with the target voice style, and the converted $F_0$ is fed to DiffVoice to convert the speech with the target voice style hierarchically. The details of each diffusion models are described as follows. 
\vspace{-0.2cm} 
\subsubsection{DiffPitch}
 
We introduce DiffPitch, a pitch generator based on the diffusion process. 
To consider continuous pitch information, we adopt a WaveNet \cite{vandenoord16_ssw} based conditional diffusion model \cite{kong2021diffwave}, which can iteratively obtain a significant receptive field with a single denoiser. The pitch encoder transforms the normalized $F_0$ of the source speech to the pitch representation $Z_{p}$. We regularized the pitch representation by pitch reconstruction loss to utilize $Z_p$ as a data-driven prior of DiffPitch as follows:
\begin{equation}
\label{pitch_loss}
  \mathcal{L}_{pitch} = \lVert X_{p} - Z_p \rVert_1.
\end{equation}
The diffusion process of DiffPitch uses the log-scale $F_0$ extracted using the YAAPT algorithm as a target ground-truth $X_{p}$.
 
The forward process of the DiffPitch is defined as follows:
\begin{equation}
\label{pitch_fwd}
    \mathrm{d} X_{p,t}=\frac{1}{2}\beta_{t}({Z_{p}} - X_{p,t}) \mathrm{d} t + \sqrt{\beta_{t}} \mathrm{d} {\mathbf{w}_{t}}\ ,
\end{equation}
where $t\in[0,1]$, $\beta_{t}$ regulates the amount of stochastic noise injected in the process, and ${\mathbf{w}_{t}}$ is the forward standard Wiener process. DiffPitch executes denoising to recover the original pitch contour in the reverse process. The reverse process of the pitch denoiser is defined as follows: 
\begin{equation}
\label{pitch_rev}
\resizebox{.98\linewidth}{!}{$
\begin{split}
    \mathrm{d}\hat{X}_{p, t} = {\small\left(\frac{1}{2}({Z_{p}}-\hat{X}_{p, t})- s_{\theta_{p}}(\hat{X}_{p,t},Z_{p}, t) \right)}\beta_{t} \mathrm{d}t +\sqrt{\beta_{t}}\mathrm{d}\bar{\mathbf{w}_{t}},
\end{split}$}
\end{equation} 
where $\bar{\mathbf{w}_{t}}$ denote the backward standard Wiener process. According to \cite{popov2022diffusionbased}, in the forward process, a sample of noisy pitch is drawn from the following distribution:\vspace{-0.25cm} 
\begin{equation}
\label{pitch_sol}
\resizebox{.98\linewidth}{!}{$
\begin{split}
    p_{t|0}(X_{p,t}|X_{p,0}) = \mathcal{N}\biggl(e^{-\frac{1}{2}\int_{0}^{t}\beta_{s}ds}X_{p,0} + &\left(1-e^{-\frac{1}{2}\int_{0}^{t}\beta_{s}ds}\right)Z_{p}\\ & ,
    \left(1-e^{-\int_{0}^{t}{\beta_{s}ds}}\right)\mathrm{I}\biggr),
\end{split}$}
\end{equation}
where $\mathrm{I}$ is the identity matrix. Distribution (\ref{pitch_sol}) is Gaussian, thus we obtain the following equation:\vspace{-0.15cm} 
\begin{equation}
\label{distribution}
\begin{aligned}
    \nabla&\log{p_{t|0}(X_{p, t}|X_{p, 0})} = \\
    &-\frac{X_{p, t}-X_{p, 0}(e^{-\frac{1}{2}\int_{0}^{t}{\beta_{s}ds}})-{Z_{p}}(1-e^{-\frac{1}{2}\int_{0}^{t}{\beta_{s}ds}})}{1 - e^{-\int_{0}^{t}{\beta_{s}ds}}}. 
\end{aligned}
\end{equation}
Therefore, DiffPitch approximates the score function with the following denoising objective: 
\begin{equation}
\label{diff_pitch_obj}
\resizebox{0.98\linewidth}{!}{$
\mathcal{L}_{p} = \mathds{E}_{X_{0}, X_{t}}\left[\lambda_{t}\big\Vert\big(s_{\theta_{p}}(X_{p,t},Z_{p},s,t)\big)-\nabla{\log{p_{t|0}(X_{p,t}|X_{p,0})}}\big\Vert_{2}^{2}\right],
 $}
\end{equation}
where $s_{\theta_{p}}$ is the pitch score estimator and $\lambda_{t}=1 - e^{-\int_{0}^{t}{\beta_{s}ds}}$.
Furthermore, we derive fast sampling using the ML-SDE solver \cite{popov2022diffusionbased}, which maximizes the log-likelihood of forward diffusion with the reverse SDE solver. During inference, the converted $F_0$ from the pitch encoder is utilized as a prior of DiffPitch, and DiffPitch generates the refined $F_0$ with the target voice style $s$. Note that we normalize $F_0$ only with the statistic of a single sentence for the fair zero-shot voice conversion scenario. 
\subsubsection{DiffVoice}
We present DiffVoice, a conditional diffusion model for high-quality speech synthesis from content, target $F_0$, and target voice style. We also utilize a data-driven prior for the diffusion models to guide the inception. According to the source-filter theory \cite{fant1970acoustic}, we first disentangle the speech components into a pitch and content representation. For a data-driven prior of DiffVoice, the source-filter encoder which consists of the source encoder $E_{src}$ and filter encoder $E_{ftr}$ reconstructs the intermediate Mel-spectrogram $Z_m$ from the disentangled speech representation as $Z_m = Z_{src} + Z_{ftr}$, where $Z_{src} = E_{src}(F_0, s)$, $Z_{ftr} = E_{ftr}(content, s)$, and $s$ denotes style embedding. Mel-spectrogram $Z_m$ is regularized as follows: 
\begin{equation}
\label{enc_loss}
  \mathcal{L}_{rec} = \lVert X_{mel} - Z_m \rVert_1,
\end{equation}
where $X_{mel}$ is the Mel-spectrogram of the ground-truth speech. Subsequently, DiffVoice can utilize the source-filter encoder output $Z_m$ as a prior, and use speaker representation $s$ as condition to maximize speaker adaptation capacity.
\begin{figure}[t]
   \centering
    \includegraphics[width=0.94\columnwidth]{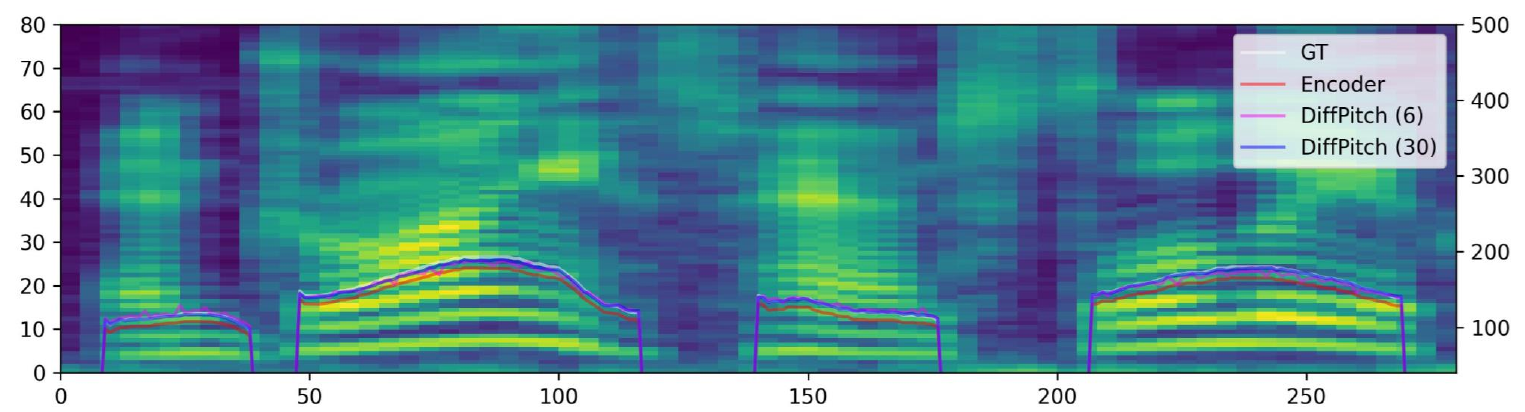}\vspace{-0.35cm}
    \caption{$F_0$ reconstruction results on $F_0$ encoder and DiffPitch}
    \label{fig2-(a)}
\end{figure}
\begin{figure}[t]
    \vspace{-0.4cm}
      \centering
    \includegraphics[width=0.92\columnwidth]{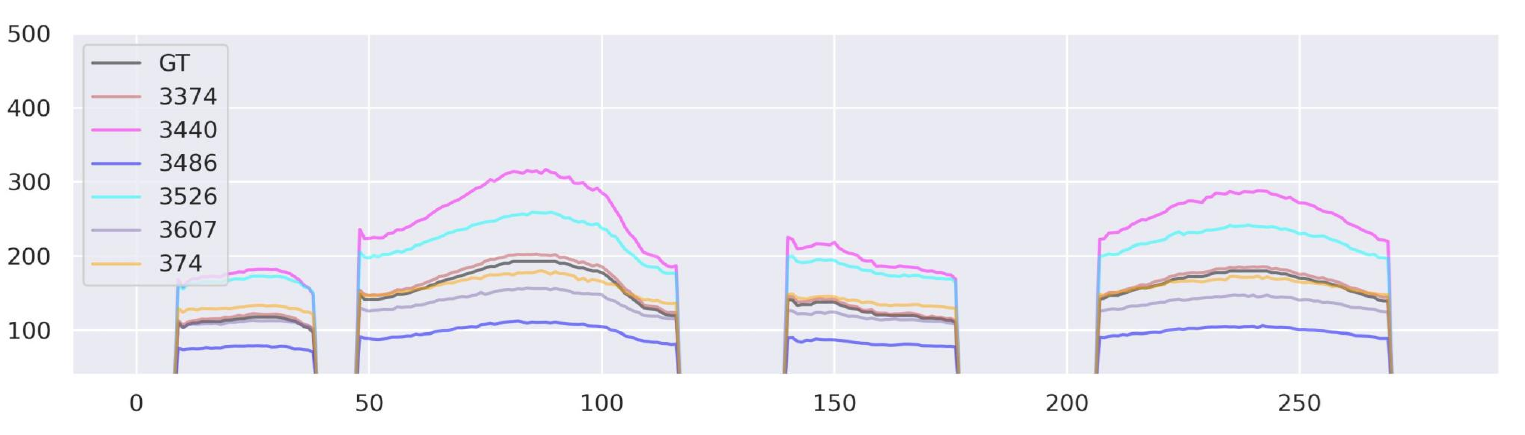}\vspace{-0.35cm}
    \caption{$F_0$ generation results of target speech}
    \label{fig2-(b)}\vspace{-0.6cm}
\end{figure}
The following equation describes the forward process of DiffVoice:
\begin{equation}
\label{mel_fwd}
\mathrm{d} X_{m,t}=\frac{1}{2}\beta_{t}({Z_{m}} - X_{m,t}) \mathrm{d} t + \sqrt{\beta_{t}} \mathrm{d} {\mathbf{w}_{t}}.
\end{equation}
The reverse process of DiffVoice is defined by:
\begin{equation}
\label{mel_rev}
\resizebox{0.98\linewidth}{!}{$ 
\begin{split}
  \mathrm{d}\hat{X}_{m, t} = {\small\left(\frac{1}{2}({Z_{m}}-\hat{X}_{m, t})- s_{\theta_{m}}(\hat{X}_{m,t},Z_{m}, t) \right)}\beta_{t} \mathrm{d}t +\sqrt{\beta_{t}}\mathrm{d}\bar{\mathbf{w}_{t}}.
\end{split}$}
\end{equation}
In the forward process, a sample of noisy Mel-spectrogram $X_{m,t}$ is taken in the same manner as equation (\ref{pitch_sol}). Finally, the objective of training the Mel-spectrogram noise estimation network $s_{\theta_{m}}$ is to optimize the score matching loss:
\begin{equation}
\label{diff_mel_obj} 
\resizebox{0.98\linewidth}{!}{$
\mathcal{L}_{m} = \mathds{E}_{X_{0}, X_{t}}\left[\lambda_{t}\big\Vert\big(s_{\theta_{m}}(X_{m,t},Z_{m},s,t)\big)-\nabla{\log{p_{t|0}(X_{m,t}|X_{m,0})}}\big\Vert_{2}^{2}\right] .
 $}
\end{equation}
During inference, the source-filter encoder takes a content representation from the source speech, target voice style $s$, and the converted $F_0$ from DiffPitch with the target voice style. The converted Mel-spectrogram  $Z_m$ from the source-filter encoder is used as a data-driven prior, and DiffVoice generates the converted speech conditioned with the target voice style.  
  
\subsection{Denoising models with masked prior}
Although the data-driven prior can significantly improve the conversion performance, DiffVoice may rely on the reconstructed Mel-spectrogram in the source-filter encoder.
To improve generalization performance of DiffVoice, we introduce a masked prior to the denoising diffusion models. 
Before fed to the DiffVoice, the prior $Z_m$ is masked, and the diffusion network jointly learns the reconstruction and denoising process. Consequently, the model can reconstruct the masked area from the surrounding context. Specifically, we apply frequency masking by interpreting continuous pitch information from a contextual point of view. 
 
\section{Experiment and result}
\subsection{Experimental setup}
\subsubsection{Datasets and preprocessing}
We train the model with a large-scale publicly available multi-speaker dataset, LibriTTS \cite{zen2019libritts}. We utilize the $\textit{train-clean-360}$ and $\textit{train-clean-100}$ subsets of LibriTTS, which contain 245 hours of speech from 1,151 speakers. We additionally use $\textit{dev-clean-other}$ subsets of LibriTTS for validation. Then, we use VCTK dataset \cite{veaux2017superseded} to evaluate the zero-shot VC performance. We randomly select sentences from the paired speech of VCTK dataset. We downsample the audio to 16 kHz, and transform the audio into a log-scale Mel-spectrogram with 80 bins using short-time Fourier transform (STFT) and Mel-filters. We use a hop size of 320 and a window size of 1,280 to map the time-resolution of the self-supervised speech representation.  

\subsubsection{Training}
We train the model using LibriTTS for 2M steps with a batch size of 64 on two NVIDIA A100 GPUs (five days), and use AdamW optimizer with the setting of \cite{lee2022hierspeech}, and implemented the learning rate schedule with a decay of $0.999^{1/8}$ at an initial learning rate of $5\times10^{-5}$. We segment the audio clip into 35,840 frame during training. For fine-tuning, we set the initial learning rate to $2\times10^{-5}$. A non-causal dilated WaveNet with 128 dimensions is used for all encoders, and DiffPitch uses the DiffWave with 64 dimensions and additional conditional layers for the pitch and style representations. DiffVoice consists of a 2D-UNet structure with the initial channel of 64 and three blocks, and the dimension of blocks are [64, 128, 256]. Following \cite{popov2022diffusionbased}, we use the noise schedule parameters of $\beta_0$ and $\beta_1$ with 0.05 and 20 respectively. The masking ratio is set to 30\% for the masked prior. For vocoder, we train the HiFi-GAN \cite{kong2020hifi} with the same training dataset, and we only replace the discriminators with multi-scale STFT discriminator of EnCodec \cite{defossez2022high}. 

\subsection{Analysis on ${F_{0}}$ prediction}
Most previous VC systems utilize normalized or quantized $F_0$ for speaker-irrelevant pitch modeling. However, we estimate a raw $F_0$ with a target voice style for better speaker adaptation. We compared three $F_0$ prediction methods: $F_0$ transformation with a statistic of $F_0$, simple $F_0$ prediction with WaveNet, a diffusion-based $F_0$ prediction with DiffPitch. Figure \ref{fig2-(a)} depicts that DiffPitch with 30 iteration steps has a similar $F_0$ contour with the ground-truth $F_0$. Figure \ref{fig2-(b)} also show the diversity of pitch contours with different target voice styles. Hence, we utilize the converted $F_0$ by DiffPitch during VC with DiffVoice. 
 
\begin{table*}[t]
    \caption{Zero-shot VC results on unseen speakers from VCTK dataset} \vspace{-0.3cm}
  \label{zvct2}
  \centering
      \resizebox{0.97\textwidth}{!}{
  \begin{tabular}{l|c|cc|cc|ccc}
    \toprule
     Method & iter. &  nMOS ($\uparrow$) & sMOS ($\uparrow$)  & CER ($\downarrow$)  & WER ($\downarrow$) & EER ($\downarrow$) &SECS ($\uparrow$)  & Params. \\
    \midrule
    GT                 & - & 3.68$\pm$0.09  & 3.59$\pm$0.03 & 0.21 & 2.17 & - & - & - \\
    GT (Mel + Vocoder) & - & 3.70$\pm$0.09  & 3.42$\pm$0.04  & 0.21 & 2.17 & - & 0.989 &13M \\
    \midrule
    AutoVC \cite{qian2019autovc} & - & 3.56$\pm$0.09 & 2.63$\pm$0.07 & 5.14 & 10.55 & 37.32 & 0.715 &30M \\
    VoiceMixer \cite{lee2021voicemixer} & - & 3.59$\pm$0.09 & 2.98$\pm$0.06 & 1.08 & 3.31 & 20.75 & 0.797  & 52M\\
    SR \cite{polyak21_interspeech} & - & 3.51$\pm$0.10 & 2.83$\pm$0.06 & 5.14 & 10.55 & 37.32 & 0.715 &15M \\
    \midrule
    DiffVC*\cite{popov2022diffusionbased}& 6 / 30 & 3.39$\pm$0.09 / 3.48$\pm$0.09 & 2.81$\pm$0.06 / 2.88$\pm$0.06   & 6.86 / 7.51 & 13.77 / 14.42 & 9.25 / 10.05 & 0.826 / 0.842 &127M\\
    DiffVC \cite{popov2022diffusionbased}& 6 / 30 & 3.63$\pm$0.09 / 3.63$\pm$0.09  & 2.98$\pm$0.06 / 2.94$\pm$0.06 & 5.82 / 6.92 & 11.76 / 13.19 & 25.30 / 24.01 & 0.786 / 0.785 &123M\\
    \textbf{Diff-HierVC (Ours)} & 6 / 30  & \textbf{3.70$\pm$0.09} / \textbf{3.74$\pm$0.09} &  3.03$\pm$0.06 /  3.02$\pm$0.06  & \textbf{0.83} / \textbf{1.19} & \textbf{3.11} / \textbf{3.58} & 3.29 /  3.66 & 0.861 / 0.860 & 18M \\
    \midrule
    \textbf{Diff-HierVC-Finetune (Ours)} & 6 / 30 & 3.65$\pm$0.09 / 3.66$\pm$0.09 & \textbf{3.04$\pm$0.05} / \textbf{3.07$\pm$0.05} & 0.97 / 1.34 & 3.15 / 3.75 & \textbf{1.50} / \textbf{1.26} & \textbf{0.894} / \textbf{0.894} &18M\\
    \bottomrule
  \end{tabular}
   }\vspace{-0.3cm} 
\end{table*} 

\subsection{Zero-shot VC}
\begin{figure}[t]
   \centering
    \includegraphics[width=1\columnwidth]{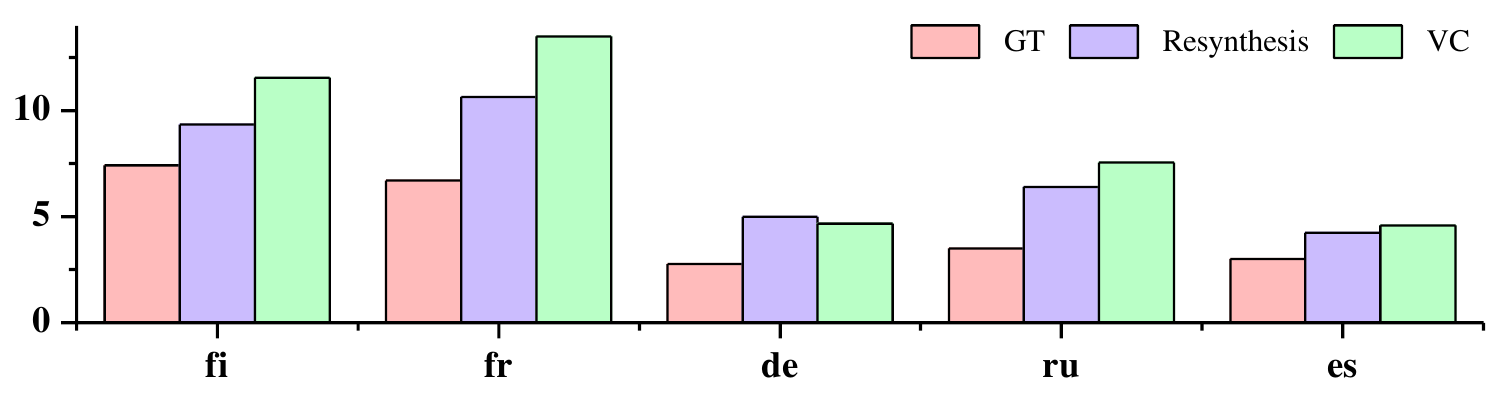}\vspace{-0.4cm}
    \caption{CER results for zero-shot cross-lingual VC on CSS10}
    \label{fig4}\vspace{-0.65cm}
\end{figure}

We conduct various subjective and objective evaluation on the zero-shot VC scenario with three models: (1) autoencoder based VC model, AutoVC \cite{qian2019autovc}, (2) GAN based VC model, VoiceMixer, (3) unit-based end-to-end speech model, Speech Resynthesis (SR) \cite{polyak21_interspeech}, and (4) diffusion-based VC model, DiffVC\footnote{To train with the same settings as our model, we trained the speaker encoder of DiffVC with train-clean-100 and 360 with 1,151 speakers. Also, DiffVC* denotes the VC results using the official checkpoint. The official code implementation uses a speaker encoder trained with a large multi-speaker dataset (voxceleb1, voxceleb2, and LibriTTS-other) containing 8,371 speakers to extract and use the speaker embedding.}. Following \cite{lee2022hierspeech}, we conduct the naturalness and similarity mean opinion score (nMOS and sMOS, respectively). Table \ref{zvct2} depicts that our model has a better nMOS and sMOS than the others. Specifically, our model achieves significantly improved content consistency\footnote{We use an automatic speech recognition model, Whisper-large \cite{radford2022robust}, and calculate the character error rate (CER) and word error rate (WER) on the 400 converted speeches with the text normalizer} and the speaker adaptation performance\footnote{For 400$\times$20 = 8,000 paired speeches, we measure the equal error rate (EER) of automatic speaker verification model \cite{kwon2021ins}. We use Resemblyzer to calculate the speaker encoder cosine similarity (SECS).}. In addition, we conducted cross-lingual VC to demonstrate zero-shot conversion performance in unseen languages. Figure \ref{fig4} shows the robust generalization performance of our model in both resynthesis and VC scenarios, even for unseen languages. Furthermore, we fine-tune the model using only one sample per speaker. Fine-tuning with small steps (1,000 steps) can improve the performance of speaker adaptation. However, the model fine-tuned with more steps shows a lower robustness of content consistency by decreasing the CER and WER. 
\vspace{-0.1cm}\subsection{Ablation study}
\label{ablation_S}
\vspace{-0.1cm}\subsubsection{Pitch modeling}
We compared three pitch modeling methods: DiffPitch, $F_0$ transformation with denormalization (Denorm.) \cite{qian2020f0}, and a simple $F_0$ prediction with the $F_0$ Encoder. All methods utilize the same normalized $F_0$ of the source speech to convert the $F_0$ with target voice style. Although Denorm. could transform the normalized $F_0$ with the mean and variance of target speech, inaccurate $F_0$ extracted from target speech decreases the voice style transfer performance regarding CER and WER with a mispronunciation and inaccurate intonation as indicated in Table \ref{ablationT}. In addition, using only the $F_0$ encoder decreases the voice style transfer performance with a higher EER than the DiffPitch even with the same WaveNet structure. 

\vspace{-0.1cm}\subsubsection{Data-driven prior and masked prior}
\label{ablation_prior}
\vspace{-0.1cm}As indicated in Table \ref{ablationT}, the data-driven prior significantly improves the voice style transfer performance. However, we found that the diffusion models may rely on the performance of source-filter encoder and the diffusion models slightly reflect the conditional information to generate the converted speech. In addition, the inaccurate ground-truth $F_0$ extracted by YAPPT is sometimes fed to the models during training and inference. Employing masked prior improves the performance with better generalization on the diffusion models taking advantage of data-driven prior.
In addition, an experiment was carried out to determine the suitable masking ratio, and as a result of Table \ref{ablation_masking}, a masking ratio of 30\% performed best. 
\vspace{-0.1cm}\subsubsection{Source-filter encoder}
\vspace{-0.1cm}It is well known that disentangling the speech plays a important role in controlling speech representation. In this work, we adopt the source-filter (SF) encoder to disentangle the speech components and regulate the starting point of diffusion models. To evaluate the effectiveness of source-filter encoder, we replace the source-filter encoder with a single encoder. Using the single encoder decreases the performance of voice conversion of the entire model, which could not appropriately disentangle the speech representation and it results in the converted speech for a prior of DiffVoice having a lower speaker similarity with the target speech as indicated in Table \ref{ablationT}. 
 
\begin{table}[t]
\centering
\caption{Results of ablation study on zero-shot VC tasks with unseen speakers from VCTK dataset. For all methods, the number of sampling iterations is 6.}\vspace{-0.25cm}
\label{ablationT}
\resizebox{1.0\columnwidth}{!}{
\begin{tabular}{l|cc|c|cc}
\toprule
 Method &  nMOS & sMOS  & CER  & EER &SECS  \\
\midrule
\textbf{Diff-HierVC}  & \textbf{3.86$\pm$0.06} & \textbf{3.02$\pm$0.09} & 0.83  & \textbf{3.29}   & \textbf{0.861} \\
Denorm. + DiffVoice & 3.81$\pm$0.06 & 3.00$\pm$0.10 & 2.67 & 5.25  &  0.850 \\
F$_{0}$ Encoder + DiffVoice    & 3.83$\pm$0.06 & 3.00$\pm$0.09 & 0.89 & 4.09 & 0.857 \\
\midrule
w.o Masked Prior   & 3.83$\pm$0.06 & 2.91$\pm$0.10 & 0.82 &  4.52  & 0.852 \\
w.o Data-driven Prior  & 3.81$\pm$0.06 & 2.90$\pm$0.10 &  0.56  & 12.77  & 0.823 \\
w.o SF Encoder   & 3.83$\pm$0.06 & 3.01$\pm$0.10 & 0.68   & 6.75  & 0.847  \\
\midrule
DiffPitch + SF Encoder & 3.77$\pm$0.06  & 2.95$\pm$0.10 & \textbf{0.30}   & 5.26  & 0.854  \\
\bottomrule
\end{tabular}
}\vspace{-0.1cm}
\end{table}

\begin{table}[t] 
\caption{Results of ablation study on different masking ratio}\vspace{-0.3cm}
  \label{ablation_masking}
  \centering
      \resizebox{0.83\columnwidth}{!}{
  \begin{tabular}{c|cccccc}
    \toprule
    Metric & 0\% & 10\% & 30\% & 50\% & 70\% & 90\% \\
    \midrule
    CER ($\downarrow$)  &  0.82  &  0.70  &  0.83  & 0.86  &  0.89  & 0.96  \\
    EER ($\downarrow$)  &  4.52  &  4.55  &   3.29   & 3.75 & 3.74 & 3.75   \\
 \bottomrule
  \end{tabular}
  }\vspace{-0.6cm}
\end{table}

\section{Conclusion}
In this paper, we presented Diff-HierVC, a diffusion-based hierarchical VC system for high-fidelity converted pitch and Mel-spectrogram generation. DiffPitch improves performance in terms of speaker similarity and phonetic intelligibility. Then, DiffVoice restores high-quality speech through a denoising process. 
Subsequently, for better generalization of the diffusion model, we proposed a masked prior that can be robustly converted by considering the context and diffusion conditions. 
Consequently, our model outperformed the state-of-the-art in all metrics even with 6.8$\times$ fewer parameters, and we demonstrated the feasibility of building the zero-shot cross-lingual VC system, which can \textit{Break Down Barriers} on various low-resource speech and language technologies. However, although our methods can significantly improve speaker adaptation quality, there are cases where the noise of input data is also considered as style. Hence, there is room for improvement towards high-quality and noise-free audio. In future work, we will decouple the noise and speech style with noise augmentation to generate high-fidelity audio even in a noisy environment.

\section{Acknowledgements}
This work was partly supported by Institute of Information \& Communications Technology Planning \& Evaluation (IITP) grant funded by the Korea government (MSIT) (No. 2019-0-00079, Artificial Intelligence Graduate School Program (Korea University) and No. 2021-0-02068, Artificial Intelligence Innovation Hub) and ESTsoft Corp., Seoul, Korea.
 
\newpage  
\bibliographystyle{IEEEtran}
\bibliography{Diff-HierVC} 
\end{document}